# The Ancient Astronomy of Easter Island: The Mamari Tablet Tells (Part 2)

Sergei Rjabchikov[1]

[1]The Sergei Rjabchikov Foundation - Research Centre for Studies of Ancient Civilisations and Cultures, Krasnodar, Russia, e-mail: srjabchikov@hotmail.com

**Abstract**

The beginning of the calendar record inscribed on the Mamari tablet has been dated to the day of the summer solstice of December 20, 1680 A.D. The moon was not visible earlier at night. Because of a possible solar eclipse it was a perilous day, a precursor of the future misfortunes: the motion of Halley's Comet of 1682 A.D. and the rebellion of the western tribes. The new data about the watchings of the star Antares have been obtained, too.

## Introduction

The civilisation of Easter Island is famous due to their numerous ceremonial platforms oriented on the sun (Mulloy 1961, 1973, 1975; Liller 1991). One can therefore presume that some folklore sources as well as *rongorongo* inscriptions retained documents of ancient priest-astronomers.

## The Mamari Calendar Text

Here the calendar text (Ca 4-9) on the Mamari board (C) is presented with its complete translation, see figure 1. During the studies of this record some clue readings have been gained. Barthel's (1958: 242-247) studies of the vast fragment of this calendar (Ca 6-9) are significant. The record reads as follows:

(1) **137a 137b 24 26** *Raa raa (tea) ari, maa*. The sun (shone) brighter (and) brighter.
(2) **6 9 5 69 9** *Ha Niva atua Moko, niva*. (It was) the watching during the dark moon Lizard (= the first night *Hiro*; the moon was absent in the sky).
(3) **6-6-6 44-9 5 69 5** *Hahaha Ta-niva atua Moko atua*. (It was) the (careful) watching during the dark moon Lizard (= the first night *Hiro*).
(4) **25 4-26 25 4-26 25 4-26 72var-72var** *Hua tuma, hua tuma, hua tuma manumanu*. The sons (= initiates) of (their) parents were as birds [= the children during the initiation rites at the ceremonial village of Orongo].
(5) **69** *Moko*. (It was) the moon Lizard (= the first night *Hiro*).
(6) **44** *Taha*. The sun set.
(7) **25 56-56 11-11** *Hua poopoo, mangomango*. Fish *poo* (and) sharks were numerous.
(8) **173 26-4 133 4** *Kupenga matua, koreha, atu*. (It was) a net from a canoe (*matua*), (there were) eels (and) fish *atu*.
(9) **25-25 56 11-11 30-30** *Huahua poo, mangomango, anaana*. Fish *poo* (and) sharks were numerous.
(10) **3 17 4-40** *Marama te Tirea* (or *Turea*). The crescent *Tireo* (= *Tueo*).
(11) **44 7 25** *Taha Tuu Hua*. The star Aldebaran (α Tauri) turned in the sky.
(12) **68 6-6** *Hono haha*. The count of the dark moons (was conducted).
(13) **108 8** *Hiri Matua*. The stars β and α Centauri lifted themselves.
(14) **44 4-6 12-12-12 44 12 4-6** *Taha tuha ika-ika-ika, taha ika tuha*. (It was) the season of catching all the fishes.
(15) **21 3 7-7 26-26-26 4-15** *Ko hina tuutuu maa-maa-maa atua roa*. The month (*Koro*, December chiefly) came when was the brightest sun.
(16) **26 139-67 139-67 139-67 4-15** *Maa taka pi, taka pi, taka pi atua roa*. The brightest fertile sun was the great god.
(17) **6-6-6 4-15** *Hoa Hoa Hoa atua roa*. The great god *Hoa* [The Master, Owner, Friend] (= *Hoa-hakananaia*).
(18) **89 45 89 28 44 116 68 4-15** *Nanaia Pua-nanaia nga taha rake, honui atua roa*. The King Going Quickly [the deity of the statue Hoa-hakananaia] of the numerous sooty terns (figuratively) was the great god.
(19) **68 3 6 3** *Hono marama a marama*. A moon was added to another moon.

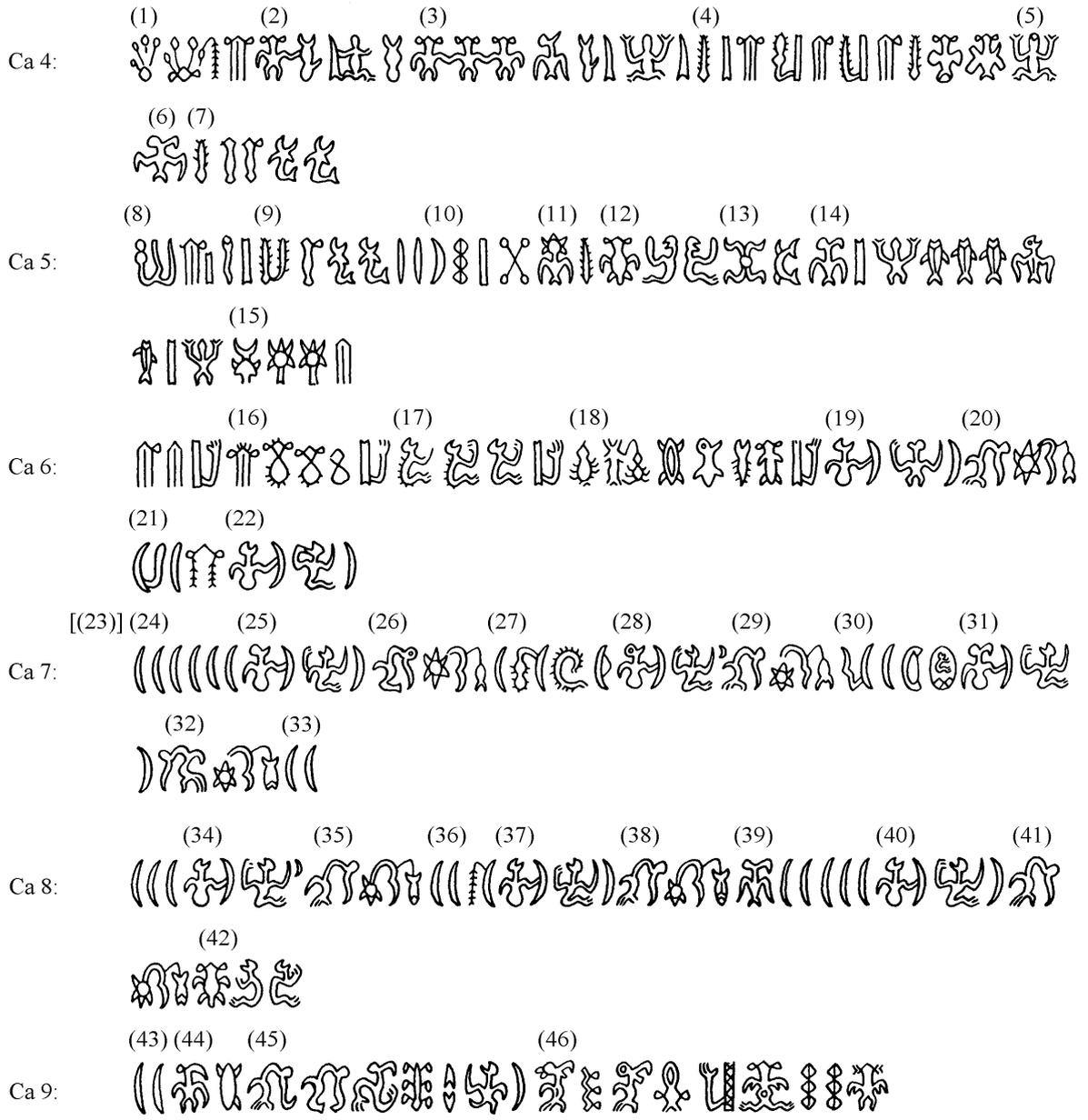

Figure 1.

(20) **19 7 50 12** *Ku tuu i Ika*: (The moons) came to the Fish (= the moon goddess):
(21) **3 4 3 24-24** *marama Ata, marama Ariari*. the moon *Ata*, the moon *Ariari* (= *Ari*).
(22) **68 3 6 3** *Hono marama a marama*. A moon was added to another moon.
[(23)] [**19 7 50** *Ku tuu i:*] [(The moons) came to:]
(24) **3 3 3 3 3 3** *marama, marama, marama, marama, marama, marama*. the six moons *Kokore*.
(25) **68 3 6 3** *Hono marama a marama*. A moon was added to another moon.
(26) **19 7 50 12** *Ku tuu i Ika*: (The moons) came to the Fish (= the moon goddess):
(27) **3 25 3 14 3** (or **3 149var**) *marama, Hua marama, Haua, marama Hotu*. the moon [*Maharu*], the moon *Hua*, (the moon) *Haua* (= *Atua*), the moon *Hotu*.
(28) **68 3 6 3** *Hono marama a marama*. A moon was added to another moon.
(29) **19 7 50 12** *Ku tuu i Ika*: (The moons) came to the Fish (= the moon goddess):
(30) **50 3 3 106 88** *(H)i marama, marama Rakau, Ma-tohi*. the moon *(H)i* (= *Ina-I-ra*, Mangarevan *Hiru* < *Hi rua*), the moon *Rakau*, the moon *Ma-tohi* (*Omotohi* = the Bearing [Moon]).
(31) **44-68 3 6 3** *Ta-hono marama a marama*. A moon was added to another moon.
(32) **19 7 50 16** *Ku tuu i Kahi*: (The moons) came to the Tuna Fish (= the god *Tangaroa*):

(33) **3 3 3 3 3** *marama, marama, marama, marama, marama*. the five moons *Kokore*.
(34) **68 3 6 3** *Hono marama a marama*. A moon was added to another moon.
(35) **19 7 50 16** *Ku tuu i Kahi*: (The moons) came to the Tuna Fish (= the god *Tangaroa*):
(36) **3 3 24 3** *marama, marama, Ai marama*. the moon [*Tapu mea*], the moon [*Matua*], the moon *Ai* [(This) place literally] = *Rongo* (*Orongo*).
(37) **68 3 6 3** *Hono marama a marama*. A moon was added to another moon.
(38) **19 7 50 16** *Ku tuu i Kahi*: (The moons) came to the Tuna Fish (= the god *Tangaroa*):
(39) **44 3 3 3 3 3** *Ta(h)a marama, marama, marama, marama, marama*. the moon *Ta(h)a* 'The Frigate Bird' = *Tane* (*Rongo Tane*), the moon [*Mauri-nui*], the moon [*Mauri-kero*], the moon [*Mutu*], the moon [*Moko = Hiro*].
(40) **68 3 6 3** *Hono marama a marama*. A moon was added to another moon.
(41) **19 7 50 16** *Ku tuu i Kahi*: (The moons) came to the Tuna Fish (= the god *Tangaroa*):
(42) **68 6-6** *hono haha*: the count of the dark moons (was conducted):
(43) **3 3** *marama, marama*, (they were) two (such) moons (at the end of the lunar month),
(44) **49-28** *maunga*. (it was) the end.
(45) **19 44b 44-25 29 6 3** *Ku tua tahua rua a marama*. During the moon [*Moko = Hiro*] the solar (*tahua*) eclipse (*tua*, *rua*) (was possible).
(46) **19 52 19 73 15 4 148 17-17 150** *Ku hiti, ku hea ro atu* "birdman"/*raa teatea*, "birdman"/*hotuhotu*. (But) the bright (and) productive sun (the bird-man literally) elevated itself (as usual).

**The Mathematical Records: Glyph 68 *hono*, *ono* 'to add, to join'**

Consider again two fragments of the previous record, see figure 2.

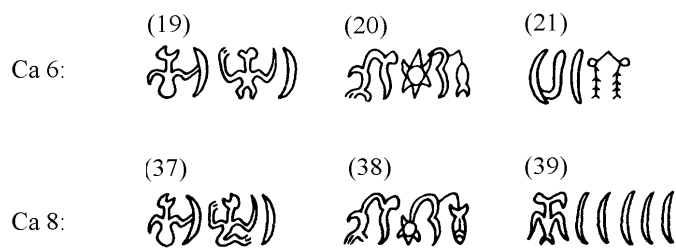

Figure 2.

Segments (19) and (37) read **68 3 6 3** *Hono marama a marama*. 'A moon was added to another moon.' Segment (20) reads **19 7 50 12** *Ku tuu i Ika*. '(The moons) came to the Fish (= the moon goddess).' Here the Fish is the symbol of the moon goddess *Hina*. According to the Rapanui myth "*Hena Naku*, the god of feathers" (Felbermayer 1971: 34-36), that goddess turned herself into a fish. In segment (38) glyph **16** is presented: **19 7 50 16** *Ku tuu i Kahi*. '(The moons) came to the Tuna Fish (= the god *Tangaroa*).' The reason is that from one to four nights of the second half of the lunar calendar, they are devoted to the deity *Tangaroa* in different Polynesian languages. In the Rapanui lexicon it is the night (moon) called *Tapume* (*Tapu mea*). Segments (21) and (39) contain the lunar names.

A lunar calendar is incised on a panel at the ceremonial platform Ahu Raai (Lee 1992: 180, figure 6.16; Rjabchikov 2014a: 3-4). There 29 crescents and 26 dots (cupules) are revealed. One can suggest that with the aid of the dots those phases of the moon were counted when it was seen clearly in the sky. The three turtle glyphs **68** *hono*, *ono* 'to add, to join' were special markers to count the lunar phases.

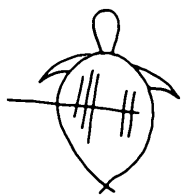

Figure 3.

I recently investigated the vast picture on another panel at Ahu Raai as a dedication to the landing of the legendary king *Hotu-Matua* (Lee 1992: 108-109, figure 4.107; the interpretation in Rjabchikov 2016a: 4). It is well known that this ruler together with his crew left the island Hiva in September (*Hora-nui*) and reached Easter Island in October (*Tangaroa-uri*) (Barthel 1978: 103, 156). In view of the fact (Rjabchikov 1993: 133-134) that the year began in the month *Maru* (*Maro*; June chiefly), the king set out on his voyage in the fourth month, and he finished it in the fifth month.

Consider now the compound design that is a fragment of that vast rock drawing, see figure 3. Inside the turtle the long line is presented. On this line two groups of short lines are plotted, namely the 3 ones and 2 ones. Using the turtle glyph **68** in order to add numbers 3 and 2, we have received number 5, the designation of the month of the arrival of *Hotu-Matua* and his people on Easter Island. Hence the interpretation of the turtle glyph **68** as the addition operator is some contexts is quite correct.

To gain a better understanding of the count of months in the local rock art consider a drawing on another panel at Ahu Raai (Lee 1992: 79, figure 4.59). Lee disclosed here the symbols of canoes, fishhooks, a shark and other sea forms. The fifth month (*Tangaroa-uri*) was the time when the fishing was allowed after several months of the strict taboo for commons (Englert 2002: 200-201). Count cupules (dots) that are gathered above a canoe on the right. These 5 dots denote the fifth month (October chiefly), the time of the fishery for all. I have recognised the fish depicted near the shark: it is a bonito (cf. Marquesan *atu* 'bonito,' Rarotongan *atu* 'ditto'). Supposedly, the valuable *atu* fish mentioned in the local folklore and in *rongorongo* records are bonitoes, too. Glyph **30** *ana* in the upper part of the picture is the symbol of the abundance and fertility.

One can now realise the sense of the parallel records about fishes inserted into the Mamari calendar record, see figure 1, segments (7), (8), (9), (14). The natives believed that the hot sun was the reason of increasing all the fishes. The other parallel records are presented on the Great St. Petersburg (P), Great Santiago (H) and Small St. Petersburg (Q) tablets (Rjabchikov 2014b: 173; 2016b: 7, figure 9).

**About an Astronomical Record on Another Board**

Consider the record on the Aruku-Kurenga (B) tablet, see figure 4.

Bv 2: 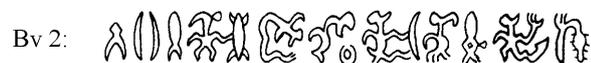

Figure 4.

Bv 2: **54 146 12 49-28 31 62 39 6 3 19 73 6 102 15-25** *Kai* 'the new moon' *ika maunga. Make too raa a marama. Ku hea a Ure Rohu.* A solar eclipse during the new moon at the end of the lunar month fell out. (The god) *Make-make* took the sun because of the moon. (But) the creator (the sun, the sun god *Tiki-Makemake*) elevated himself (as usual).

It is the description of a solar eclipse. This inscription has many common features with segments (44) – (46) of figure 1 (the Mamari tablet). Glyph **146** denotes 'the new moon' (the waxing moon because of the waning moon); by the way, the similar glyph **140** denotes 'the full moon' (the waning moon because of the waxing moon).

A parallel record in presented on a panel at the ceremonial platform Ahu Marate (Lee 1992: 51, figure 4.9; the interpretation in Rjabchikov 1994: 31-32, table 1; 1997a).

First of all, this place name, *Marate*, reads *Mara tea* (The white or clear Moon). A face (*aringa, mata*) is depicted on the left. A hand holding the sun (the small round with the dot in its centre) is shown on the right of the picture. I think it is the designation of a solar eclipse. Here different groups of dots (cupules) divided by different symbols (wings, birds) – 23 – 17 – 7 – 18 – 13 – are incised. The first three numbers check nicely with the three numbers (23/16/7) on a calendar device manufactured in Siberia in the Palaeolithic times. Thus, the Rapanui priest-astronomers could predict solar and lunar eclipses. It is of interest that the picture stretches near a water reservoir. Not surprisingly, I have remembered such a place name without the precise position on local maps: *Vai Tataku Po* (Barthel 1978: 264, 268). It means 'The water reservoir (is beside the spot where somebody) counts nights.'

## Some Ethnographical Aspects of the Festival of Initiations

Before the month *Koro* (December chiefly) the children of the king and of the aristocracy lived during three months together with some priests and teachers in caves on the islet Motu Nui to prepare themselves for the initiation rituals (Métraux 1940: 105). I suggest that the isolation extended from the middle of the month *Hora-nui* (September chiefly) to the middle of the month *Koro* (December chiefly). A local song dedicated to these ceremonies sounded as follows:

*Kia Mahe Renga te hiva te manu ko te hiva katoo no koe ehuru oke a umu ko marie he manu haka ohiohi o Mahe Renga a tapua ara tahe o te iva* (Routledge 1914-1915).

My own research has given this reconstruction of the chant:

*Kia Ma-he-Renga te Hiva, te manu ko te Hiva. Ka too no koe e huru. Oke a umu ko mari e he manu. Hakaohiohi (= hakaoioi) o Ma-he-Renga a tapua ara tahe (= taha) o te Iva (= Hiva).* (It is the speech) to *Ma-he-Renga* (the name of a boy who participated in the bird rites on Motu Nui) from Hiva (the legendary homeland), a bird from Hiva is speaking (*kia*). (Oh child,) take feathers! Make the meal from the earth oven (dedicated) to eggs (*mari*, *mamari*) of birds! *Ma-he-Renga* (as the bird) is moving along the sacred area (*tapua*, *tapu*) of the road (*ara*) of the frigate birds (*taha*; sooty terns figuratively) from Hiva.

The initiation rites (*take*) were performed over the children at the sacred village of Orongo on the island. The bodies were painted red and white. The rounds were depicted on the backs and backsides (Métraux 1940: 105). A local song called *take* (initiation) sounded as follows:

*Katuu mai e te take na kahu par ravarava take koai to tua angakope komata mahore apero ta a mee o korua e aka-aka no ena e mitimiti ena* (Routledge 1914-1915).

My own research has given this reconstruction of the chant:

*Ka tuu mai e te take na kahu. Pa ravarava take ko ai to tua anga kope. Ko mata ma hore a pero ta a mee o korua e akaaka no ena e mitimiti ena.* The clothes (*kahu*) were taken off (*tuu mai*) for the *take* rite. There were places (*ko ai*) of the signs *take* (*pa* 'sign,' *rava* 'to receive; to have,' *taka* 'round; the sun') made (*anga*) on the backs (and) back-sides (*to tua*) of boys (*kope*). The faces were for the tattoos (*hore*) in such a manner (*a pero = pera*) that the tattoos (*ta mee*) of a great number (*akaaka*, *rakerake*) of the food (*korua*) were because of the heat (*mitimiti*).

I compared earlier Old Rapanui *korua* (glyphs **21 29** on the Santiago staff, I 3) with Margarevan *akakorukoru* 'to fill the mouth up entirely with food' (Rjabchikov 1997b). The tattoos on the faces were, in my opinion, series of large facial dots known as *humu* (cf. Métraux 1940: 241). Possibly the wordplay took place: cf. Rapanui *umu* 'earth oven' (the ideas about the bright sun, heat etc.). In this text magic formulae of fertility and abundabce were encoded.

In the record on the Santiago staff (I) the ceremony of the tattooing of the rounds *take* during the initiation (cf. Rapanui *taka* 'circle') is described, see figure 5.

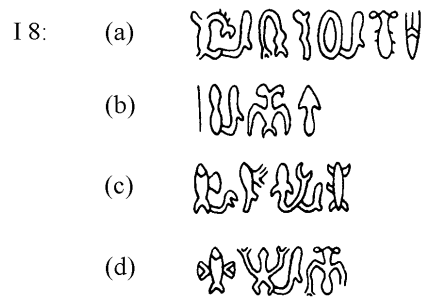

Figure 5.

I 8: (a) **62 (102) 54 105 115 (102) 56 1** (b) **(a vertical line) 65 (102) 44 22** (c) **111 (102 102) 15 11 (102) 28** (d) **111 6 (102) 68 [72]** (a) *To kai moa taka (take) po Tiki.* (b) *Rangi: Taha, ao.* (c) *Ngu: Roa Mango-*[*NGO*] [the last syllable is fixed to the word]. (d) *Ngu: a (h)ono [manu].* (a) The tattoos (*to kai*) of the cocks (= the boys metaphorically) were rounds (*take*, *taka*) as (the symbols of) the keeping of (the sun deity) *Tiki*. (b) The shout (of a priest) was: the Frigate Bird of the ceremonial paddle *ao* (the figures on the back of the statue Hoa-Hakananaia). (c) The recitation was: the Shark (the symbolism of the god *Tangaroa* or *Tini-rau*) appeared. (d) The recitation was: [The birds] were added (united, gathered)… (Cf. the Rapanui place name *Hare Moa Tataka* 'The House of Initiations.')

The last glyph **72** (or **81**) *manu* (bird) is omitted in Barthel's (1958) publication of the *rongorongo* corpus, but it is presented in Fischer's (1997) notation of the inscription.

### The Statue Hoa-Hakananaia

In compliance with Heyerdahl (1976), this monument represents the principal god *Makemake*. When the statue was taken from Orongo, the natives put the small stone idol of *Makemake* instead of it into the same house (Van Tilburg 2014).

Consider another record on the Santiago staff, see figure 6.

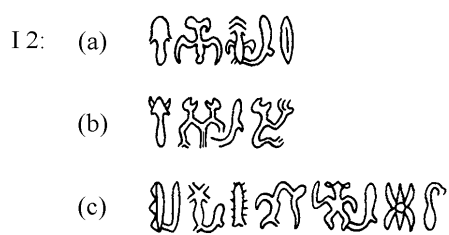

Figure 6.

I 2: (a) **22 44 68 33 (102) 110** (b) **21 6-6-(102)-6** (c) **18-4 56 (102) 48 44b 6-(102)-7 25** (a) *Ao (or rapa), Taha, Hono Vai, (Ko)mari* (b) *ko Maa Ho(a), Ho(a), Ho(a).* (c) *Te atua Pua hua, tua Hatu hua.* '(a) (The words) 'the paddles', 'the (frigate) birds', 'addition of the water' (the connection with the paddles), 'the vulvas' (b) (are inscribed on the back of the statue of) the Brightness (*Maa = Makemake*) – the great *Hoa* ('The Master, Owner, Friend' = the stone figure *Hoa-hakananaia*; the deity *Rarai-a-Hoa* or *Rarai-a-Hova = Tiki-Makemake*, *Tane*). (c) (He is) the god 'The Top' [Source; the (first) Egg etc.] giving fruits (eggs). (It is) the back of (the god) *Tiki-te-Hatu* giving fruits (eggs).' (Rjabchikov 2009a: figure 22).

It is the description of signs carved on the back of the statue Hoa-Hakananaia. The term *pua* has several meanings in the Polynesian languages (origin; source; tribe; king; chief; summit; peak etc.), cf., e.g., Rapanui *puapua* 'top, summit' (< \**pua*) and Hawaiian *puu* 'peak.' Because of the alternation of the sounds *p/h*, Old Rapanui *pua* and *hua* mean 'flower; fruit; egg; son.'

Let us study the symbols on the back of that majestic monument (Heyerdahl 1976: plate 5b). One small paddle (*rapa*) and two large paddles (*ao*) are seen in the upper part of the picture. Between the paddles *ao* the sign of the sooty tern and glyph **64** *mea* are rendered. It is the symbolism of the sacred bird known as *Manu Mea* (The Red Bird). Glyphs **1** *Tiki* (the name of the solar god), *tiko* 'menstruation,' *tika* 'straight' (*tiktia*), 'power' (*titikanga*), *komari* 'vulva' are presented on the right.

The second row of symbols contains two bird-men who look at each other. Glyph **3** *Hina* is seen on the right under the second bird-man.

The third row of symbols includes the sun (*raa*, *taka*) at noon, three levels of the heavens, and the sun god *Tane*. Two small rounds (the sun designs, *raa*, *taka*) near the latter symbol could be late carvings.

My interpretation of this intricate plot is as follows. It was the sacramental manifestation of the god-creator *Tiki-te-Hatu = Makemake*. The ceremonial paddles were signs of authority; besides, they were certain symbols of the fertility and abundance (cf. Heyerdahl 1976: figure 9). The central idea of that picture was the appearance of sooty terns (the sun signs; the symbolism of the growing warmth of the day). The moon goddess *Hina* was close related to bird-men. Two such persons sitting nearby denoted the resumption of the same rite (feast) each year.

In the real ceremonies of the election of a bird-man and of initiations some of participants could wear masks of birds and hold in their hands ceremonial paddles (*ao*, *rapa*).

### The Supplementary Data about this Monument

A Rapanui chant has been known in several records, here is one of them (Blixen 1979: 52-53):

| | |
|---|---|
| *E Pua, e Pua te Oheohe,* | The great Ruler with Paddles, |
| *E Pua te nanaia,* | The Ruler who is going quickly, |
| *E Tama te raa,* | The Child of the sun, |
| *Hiro rangi pakupaku,* | (the god) *Hiro* in the dry sky, |
| … | ... |
| | (The translation is of mine.) |

Rapanui *ohe* 'bamboo' is not suitable in the context, cf. Mangarevan *ohe* 'oar; paddle' and *hoe* 'paddle,' Rapanui *hoe* 'ditto.' This ruler is the god (statue) *Hoa-Hakananaia* (*Tiki*, *Makemake*). The big ring in the third row of the drawing on the statue's back is *Tama te raa* (The Child of the sun). The sign of the sky and the sign of the sun god *Tane* depicted beneath it in the same picture are *Hiro rangi pakupaku* (the rain deity *Hiro* in the sky in summer), cf. Maori *pakupaku*, Rapanui *paka* 'dry.'

Consider the record on the Small St. Petersburg tablet, see figure 7.

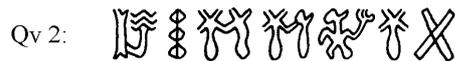

Figure 7.

Qv 2: **4-4-33 17 45-45-45 27 6-45 41** *Atuaatua/ua te Pua-Pua-Pua ro(h)u hopu are (ere).* The great god *Pua-Pua-Pua* (the sublime *Pua*) gave a servant (*hopu*) an egg (*are*, *ere*).

The natives believed that the statue Hoa-Hakananaia (the image of the god *Tiki-Makemake*) had the great supernatural power (*mana*).

Consider the record on the Great Santiago tablet, see figure 8.

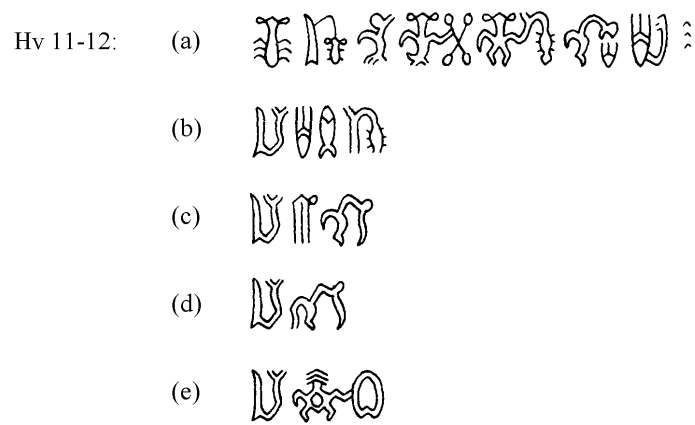

Figure 8.

Hv 11-12: (a) **89 5-89 19 6-40 6-25 62 1-1 15 24** *Nanaia atua Nanaia ki hare-ahu toa Tikitiki roa ai*: (b) **5-15 1 12 15-25** *atua roa Tiki Ika rohu*, (c) **5-15 26 44b** *atua roa Maa Tua*, (d) **5-15 44b** *atua roa Tua*, (e) **5-15 6-33 47** *atua roa Hau (H)ava*. (a) The god *Nanaia* (the elected bird-man) is going quickly to the house (served) as a platform of the warrior 'The great *Tikitiki*' of (this) place [Orongo]: (b) the great god *Tiki* (son of) the Fish (*Tangaroa*)-the crea-tor, (son of), (c) the great god *Maa* [i.e. *Makemake*] (son) of the Deep Water, (son of) (d) the great god The Deep Water, (son of) (e) (the legendary homeland) Havaiki.

Here the myth about the god *Tikitiki* (the great *Tiki*) associated with the statue Hoa-Hakananaia is presented. The Polynesians, as many others, saw their history through the prism of their ideology and mythology. The god *Tiki* was son of the sea god *Tangaroa* as it was written in Manuscript E (Barthel 1978: 304). *Maa* was the god *Makemake*. Again, as an equivalent of *Tiki*, he was son of the Deep Water (a metaphor for *Tangaroa*). The Polynesians thought that they came from the legendary homeland Havaiki. According to Taumoefolau (1996), that term sounded as PPN *Sau 'ariki* in the remote past.

The cult of the paramount god *Tangaroa* was widespread among the western tribes of Easter Island (Tuu, Moko, Hanau Momoko, Miru and others). At the New Brunkswick Museum (Saint John) a figurine manufactured from the barkcloth is housed (Kjellgren 2001: 61, figure 26). It resembles a seal. Its name could be *Niuhi* (not *Nuihi*) or *Pakia*, or even *Tangaroa*. On this puppet the symbol of the wave (cf. glyph **33** *vai*, *ua*) is depicted. This figurine indubitably had the protective functions.

Consider the parallel records on the Small Washington (R) and Aruku-Kurenga boards, see figure 9. They inform about the statue Hoa-hakananaia at the sacred village of Orongo.

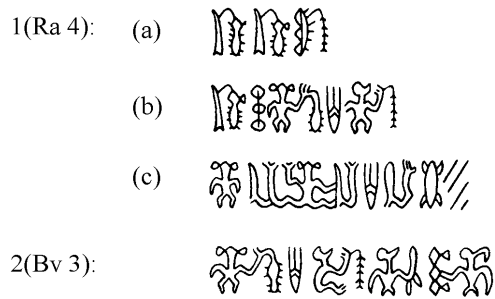

Figure 9.

1 (Ra 4): (a) **5-89-5-89 18 24** *Atua Nanaia, atua Nanaia te ai,* The god Moving Swiftly (= the statue Hoa-Hakananaia; the god *Tiki-Makemake*), the god Moving Swiftly of (this) place [Orongo],
(b) **5-89 17 6-25 1 6 24** *atua Nanaia te ahu Tiki a ai.* the god Moving Swiftly (is located) at the house (platform literally) of this place [Orongo].
(c) **6 5-15 69 5-15 1 48-15 28** (a damaged segment) *A atua roa Moko, atua roa Tiki uri Nga…* (The drawings on this statue are): the great god *Hiro*, the great god *Tiki* originated from the Egg… (Cf. Maori *nganga* 'shell; husk.')
2 (Bv 3): **6-25 1 6 24 44-51 17 44** *Ahu Tiki a ai: take te taha.* The house (platform literally) of the god *Tiki* of this place [Orongo]: (it is associated with) the initiations of the young men (or men looking like birds = in masks).

Compare segment (b) of the first record and the second record. They both contain the common segment **6-25 1 6 24**. In segment (b) glyphs **5-89** and glyph **17** precede that segment. We know that glyphs **17** and **18** are allographs (variants). Segment (a) contains the doubling of glyphs **5-89**, glyph **18** = **17**, and glyph **24**. From these consequences (and from other ones) it is apparent that glyph **17** (**18**) is a grammatical article. I have chosen these records to demonstrate the beginning of my decipherment.

It should be noted that Mangarevan *taha* means 'young.'
Consider another record on the Aruku-Kurenga tablet, see figure 10.

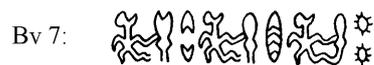

Figure 10.

Bv 7: **6-73 29 6-73 33 6-73 141-141** *Hoe rua, hoe vai* (or *ua*), *hoe takataka.* The paddle of the sunset, the paddle of the water (rain), the red paddle. (Cf. Rapanui *taka* 'ruddy,' *mataka* 'red'.)

Whereas Heyerdahl (1976) interpreted the designs of ceremonial paddles depicted at Orongo as the Old Andean weeping eye motif, Fedorova (1981) interpreted the semantics of such paddles as the pure Polynesian invention: in her opinion, they were images of the chthonic and rain god *Hiro*.

During burial rites priests were covered with black ashes and carried a ceremonial paddle *rapa* (Routledge 1998: 229). At the Smithsonian Institution (Washington) a ceremonial paddle *ao* is housed (Heyerdahl 1976: colour photo 15). It was painted red and white.

Supposedly, in the record both types of paddles are described. They were certain signs of death and resurrection in the new life (the transition from winter to summer though spring). So, the paddles *rapa* and *ao* could symbolise the death of nature (autumn, winter, evening, night, coldness, black things etc.) and its recovery (spring, summer, morning, day, warmth, heat, red and white things, etc.) respectively.

The Rapanui phrase *puku hanga oao* means 'east,' otherwise 'the rock (where) the paddle *ao* or the sunlight is moving' [*puku hanga o ao*], cf. Maori *anga* 'to move in a certain direction' and *ao* 'daytime.'

The Rapanui folklore text "*Apai*" (Thomson 1891: 517-518) contains the following segment:

*Maru matai ... aku Hoa-Hoa tae kote, kura.* The first month *Maru* (or *Maro*; June chiefly; the month of the winter solstice)... the deity (*akuaku*) *Hoa-Hoa* (i.e. the statue Hoa-Hakananaia = the god *Tiki-Makemake*) is without the white (*kote*, *kotea*) (and) red (*kura*) colours. (The translation is of mine.)

Consider another record on the Aruku-Kurenga tablet, see figure 11.

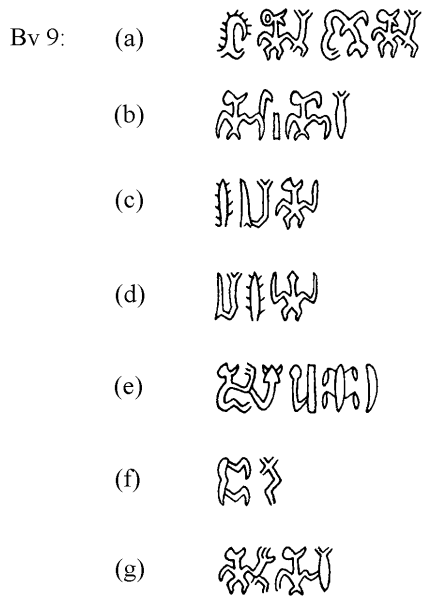

Figure 11.

Bv 9: (a) **14 19-15 31 19-15** *Haua kura, Make kura*. (The moon goddess) *Haua* is red, (the sun god) *Makemake* is red.
(b) **44 35 4 44 9** *Ta(h)a PA atua, taniva*. Frigate birds (sooty terns figuratively) SIGN (the determinative) as the deities, the darkness (the threat of sharks to servants *hopu*).
(c) **25 5-15 6** *Hua atua roa Hoa*. The egg of the great god *Hoa*.
(d) **5-15 25 6** *Atua roa Hua Hoa*. The great god 'The Egg of *Hoa*.'
(e) **6-21 4-4 25 3** *Hakatiti hua Hina*. The eggs of the moon (the moon goddess *Hina*) are numerous.
(f) **50 29 70** *Hi: rua, pua*. The rays of the sun (were visible) during the settings (and) the risings.
(g) **6-15 44 9** *Hora. Taniva*. (The season) *Hora* (the months *Hora-iti*, *Hora-nui* = August-September chiefly). The darkness (the sea waters) [as the main threat].

The text tells of the appearance of sooty terns at the beginning of September (the month of the vernal equinox). Here the red colour denotes spring-time. Rapanui *hakatiti* means 'to accumulate; to glut; to augment; to supply; to multiply; to fructify.'

One can state with assurance that main rites during the bird-man festival were conducted inside the house where the statue Hoa-Hakananaia stood.

Consider the record on the Tahua (A) tablet, see figure 12.

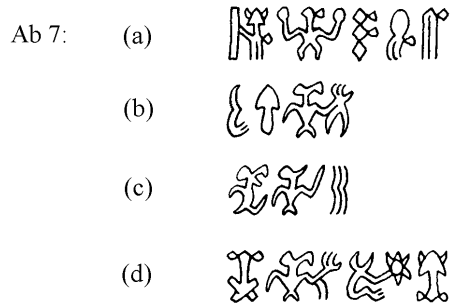

Figure 12.

Ab 7: (a) **4-26/21 6-4 17 43-73 26** *Timo ako: Hatu te Maea Maa*. The pupils are carving (writing) [the text in the school]: The stone (= statue) of the Bright Light (= the god *Makemake*) appears (= is widely available).
(b) **43 22 44 84** *Ma ao Taha ivi*. The ceremonial paddles of the forefather(s) 'The Frigate Bird(s)' ('The Sooty Tern(s)' figuratively) go.
(c) **19-44 33** *Kita vai* (or *ua*). The water in the sea is the cause of death (because of sharks).
(d) **49 44 15 6-7 21** *Mau Taha* (*Manu-tara* figuratively) *roa – Hatu oko*. The bird-man (as the image of the god *Makemake = Tiki-te-Hatu*) is taking the egg (the ripe object literally).

Here the ceremonial paddles are the symbols of the land of the primordial ancestors. It is apparent that such objects were of great concern in the cult of bird-man.

Barthel (1978: 198-199) could not translate the word *kikita* written in Manuscript E; Fedorova (1988: 82; 1993: 53) compared it with Tongan *kita* 'to fall ill,' and *fakakita* 'to give death.' Estella (1921: 131) published an ancient Rapanui text. Here *Maurata*, grandson of king *Nga Ara*, is mentioned. According to Routledge (1914-1915), *Maurata* was a servant *hopu* of the future bird-man *Utu-Piro* in the search of the first egg of sooty terns. I have translated the text. The islet Motu Nui is called *Motu manu* 'The islet of birds.' The record begins with the words: *Vai e vai e uri* 'The water (of the sea) was black.' That colour denotes the threat of death because of dangerous sharks. The term *pi* (the fullness) at the end of the text implies the abundance and fertility.

Per Routledge (1988: 262), the term *hopu* (servant) is connected with Rapanui *hopu* 'to wash.' Hence, *hopu* = 'diver; swimmer.' But another etymology is also available: cf. Marquesan *hopu* 'to seize in the arms' and Mangarevan *akahopu* 'to keep the body bent on the march,' and if so *hopu* = 'invader.'

**The Dating of the Calendar Text on the Mamari Board**

In that record the summer solstice of December 20, 1680 A.D. was described, see figure 1, segment (1). Here and everywhere else, I use the computer program RedShift Multimedia Astronomy (Maris Multimedia, San Rafael, USA) to look at the heavens above Easter Island.

The moon was invisible in the sky in that night, therefore the local priest-astronomers chose that night/day as the new moon *Hiro* (cf. the name *Moko* 'Lizard' in the text). In compliance with the record, the annul rituals – initiations of the children of the eastern tribes (Hanau Eepe, Tupa, Tupa-Hotu, Hotu, known as Hotu-Iti later) – began at the religious centre of Orongo on that day. Some rites were conducted within the house where the famous statue Hoa-hakananaia stood (now the statue stands at the British Museum, London). The day *Tireo* (*Tueo = Tureo*) also had the peculiar position in the inscription. The other names of crescents were collected together beginning with the third lunar phase.

So, the night *Moko* (*Hiro*) = December 20 (chiefly), 1680 A.D., the night *Omotohi* (*Ma-tohi*) = January 5 (chiefly) 1681 A.D. [the full moon], the night *Rongo* = January 13 (chiefly), 1681 A.D., the night *Tane* (*Rongo Tane*) = January 14 (chiefly), 1681 A.D., then the nights passed from January 15 (chiefly), 1681 A.D. to January 18 (chiefly), 1681 A.D. (the next new moon).

The night/day (twenty four hours) lasted from the twilight till the next twilight. Hence, the night/day *Moko* (*Hiro*) in our case lasted from 20:44 on December 19, 1680 A.D. till 20:44 on December 20, 1680 A.D., the night/day *Tireo* lasted from 20:44 on December 20, 1680 A.D. till 20:45 on December 21, 1680 A.D., and so on.

According to the computer model, the transit of Aldebaran was at 22:33 (December 20), otherwise in the night *Tireo*. The rising of β Centauri was at 21:56 (December 20), the rising of α Centauri was at 22:21 (December 20), otherwise in the same night. Thus, watching the elevation of α Centauri higher and higher, the priest-astronomers saw how Aldebaran drooped lower and lower.

### The Old Peruvians on Easter Island

We need not doubt that the Old Peruvians (the ancient people of the Andean area wider) visited Easter Island. The Native American-Rapanui (Polynesian) contact happened on the data of biology in 1280 – 1495 A.D. (Moreno-Mayar et al. 2014). Only a few Old Andean men could settle on the island, and soon their descendants (Hanau Eepe, Tupa-Hotu, Tupa, Hotu, later Hotu-Iti, or the Small Hotu) forgot their native language, and only some such words were retained in the late Old Rapanui language. Such processes took place in the history of many peoples.

### The Substratum: the Hurrians and the Early Indo-Europeans

The ancient state Mitanni extended from the Tigris river to the Mediterranean Sea several thousand years ago. The inhabitants of the country generally spoke the Hurrian language. The kings forgot their native Indo-Aryan language; nevertheless, in the local inscriptions (and in the speech also) only several former divine names and special terms were preserved (Wilhelm 1989).

### The Substratum: the Old Russians and the Vikings

The legendary prince *Ryurik* (the Scandinavian *Hroerekr*) was the founder of the Old Russian dynasty. In accordance with the Russian chronicles, he arrived at the Russian land together with the men *Sineus* and *Truvar* – from the Old Norse language those names mean – together with own relatives (*sine use*) and true army (*tru var*) indeed (Danilevsky 1999: 63ff).

One of the earliest runic records discovered in the Old Russian town Ladoga (Staraya Ladoga) is found on a wooden stick (Krause 1960; Melnikova 2001: 206). My own interpretation of this inscription (letter) written in Old Norse is as follows:

*Dó yfir of vaRiþR Hali ræs.* King (*ræs = ræsir*) *Hali* died at (this) spot.
*Fran. Manna grand. Fimbul sinni plóga.* (He was your) relative (*fran = frændi*). (He) did harm to men. (He was) the mighty host of ploughs (i.e. of the vast territory inhabited by peasants).

On the reverse of the stick three runes *u* called *ur* (bull) were inscribed. Presumably, it was the magic formula pertained to burial rituals.

*Hali* (*Helgi*) was the Russian prince *Oleg* (*Veshchy Oleg*). He died in 912 A.D. in accordance with Old Russian chronicles.

The alternation of the sounds *a/e* is seen in several Old Norse words: cf. *varði*, *verja* 'to defend,' *draumr* 'dream,' *dreyma* 'to dream,' and *langr*, *lengi* 'long.' A number of Russian (Old Russian) words could come from Old Norse: cf. Russian *hrabry* 'brave' and Old Norse *hrapa* 'to rush' (cf. Russian *delat' nahrapom* 'to have the insolence to do something' also), Russian *vedat'* 'to know' and Old Norse *vita* 'ditto,' Russian *knyaz'* 'prince' and Old Norse *konungr* 'king,' Russian *mech* 'sword' and Old Norse *mækit* 'ditto,' Russian *druzhina* 'army' and Old Norse *drengr* 'warrior,' *þrúðugr* 'strong,' Russian *lyudi*

'people' and Old Norse *lýðr* 'men; people,' *lið* 'people,' Russian *plug* 'plough' and Old Norse *plóga* 'ditto,' Russian *grozit'* 'to threaten,' *ugroza* 'threat' and Old Norse *grand* 'harm.'

Soon the Old Russian rulers forgot Old Norse, and spoke Old Russian (Slavonic).

### The Substratum: the Old Andean people on Easter Island

Imbelloni (1926, 1928) and Palavecino (1926) published a diversity of Old Peruvian-Polynesian lexical parallels. In my view, the comparisons of Quechua *k(h)umar(a)* 'sweet potatoes' and Rapanui *kumara* 'ditto' as well as Aymara *tayna* 'brother; sister' and Rapanui *taina* 'ditto' are absolutely reliable.

According to Knoche (1912a: 873), the Hanau Eepe ("Long Ears" in the text) erected statues on Easter Island, and the Hanau Momoko ("Short Ears" in the text) built platforms.

Barthel (1962: 847) could not translate the phrase *ko piti ko pata* in the Rapanui chant "*I Anakena au i mate ai.*" The text tells of *rano* (crater with a lake), and there are only three such craters on Easter Island: Rano Aroi, Rano Kau (or Kao) and Rano Raraku. The quarry where most of the Rapanui monuments were manufactured is at the last spot.

I have imagined this scene in the Rapanui history. A leader of the Hanau Momoko visited time and again Rano Raraku, and heard the command: "Cut a platform!" In Quechua it sounded: *P'itiy patata*! (cf. Quechua *p'itiy* 'to cut' and *pata* 'platform'). Since the reduplicated and single lexical forms often are equal in the Rapanui language, the word *patata* was preserved in the folk memory as *pata*. The grammatical articles *ko* were inserted into the obscure text later: *ko piti ko pata*. Of cource, these words can be translated as the pure Polynesian ones as well.

### A Great Literary Masterpiece on the Mamari Tablet

A drama worthy of Shakespeare is put on the Mamari board down. It is, without a doubt, a masterpiece of world literature. Just as in the First part of King Henry the Sixth, the author of the Rapanui text first appealed to the celestial deity *Tiki*, as to a mighty protector (Ca 1-Ca 2), and then told of the loss (Ca 13-Ca 14-Cb 1, Cb 12, Cb 10): the deaths of king *Aringa* (*Tupa-Aringa-Anga*) and a lot of warriors of the tribe Tupa (Tupa-Hotu, Hanau Eepe) occurred. These records were deciphered before (Rjabchikov 2012a: 567-569, figure 6, fragment 2; figure 7, fragment 2; 2012b: 17-18, figure 10; 2016c: 8, figures 10 and 11).

The Mamari inscription calls three ghosts who could defend the Tupa-Hotu tribe, see figure 13.

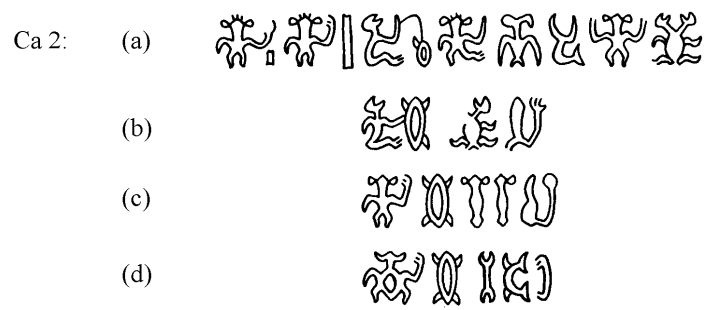

Figure 13.

Ca 2: (a) **6-4-6-4 6-28 6 44 2 6 20** *Hatuhatu, hanga a Taha Hina a Pikea (Ungu).* The moon goddess *Hina* as the Crab appeared, moved in the night called *Tane* (*Rongo-Tane*).
(b) **6-28 20 48-15** *Hanga Pikea (Ungu) Uri.* (The moon goddess) *Pikea Uri* (The Black Crab) moved.
(c) **6-28 56-56-25-4** *Hanga Popohutu.* The Surf moved.
(d) **50 6-28 84 8 14** *Hi, hanga ivi Matua roa.* The ancestor *Matua roa* (The Great Parent) appeared, moved.

(1) In conformity with some of Rapanui traditions (Blixen 1974: 7-12; Brown 1996: 222-224; Knoche 1912b: 65-67; Métraux 1940: 365-366), an old woman called *Nuahine Pikea Uri* transformed herself in some instances into a ghost looking like a black crab (*pikea uri*).

This crab (the moon goddess in fact) sometimes hid in rocks (among stones). In my opinion, the wordplay was possible: cf. Rapanui *pikea* 'crab' and *piko* 'to hide.'

According to the Rapanui legend "*Hetereki*" (Felbermayer 1971: 65-79), the witch *Nuahine Pike Uri* of the Tupa-Hotu tribe during a battle with the Miru tribe (Hanau Momoko, Ko Tuu) prayed about the victory. That priestess was named after the strong goddess premeditatedly. She became omnipotent, as the locals thought.

I conceive that the image of the crab as a protective being was an echo of the Old Peruvian culture. In the Incan beliefs the crab was the waning moon and the terrible aspect of the Great Mother (Cooper 1992: 63). On a mummy of a Moche woman (priestess, queen) different tattoos were disclosed: "stylized catfish, spiders, crabs, felines, snakes, and a supernatural being commonly called the Moon Animal" (Lobell and Powell 2013: 44). I presume that all they had the protective functions. Furthermore, an anthropomorphised crab ornament decorated the Moche tomb of the Old Lord of Sipán (Scher 2010: 90). The safety aim of that pattern is obvious.

(2) According to the Rapanui myth "*Makemake*, the protector and helper" (Felbermayer 1971: 41-44), the big waves (*te vave nunui*) associated with the deity *Makemake* played the role of a certain deity. The Rapanui god *Popohutu* ('The Surf; the Wave,' cf. Maori *pōhutu* 'surf') also was an echo of the Old Peruvian culture. Initially it was the Incan storm god *Paryaqaqa*, an aspect of the rain god; he emerged from five eggs of a falcon, and then he existed like a man (Taylor 2008: 11ff). The name *Parya Qaqa* means 'The Rain – the Rock,' cf. Quechua *para* 'rain' and *qaqa* 'rock.'

(3) The god *Matua roa* (The great Parent) is the junction of the Rapanui sun god *Atua Metua* (*Tane*) and the Old Peruvian sun god *Inti* and the god-creator *Qun Tiksi Wiraquocha*. Here it must be underscored that *Makemake* was not the Polynesian god at first, and his pure solar interpretation (*Maa-ke*, *Maa-ki*, *Maa*) was added later. According to Rapanui legend "The Fight between *Hetereki* and *Taereka*" (Métraux 1940: 379-381), a priest of the Tupa-Hotu tribe during a battle with the Miru tribe (Hanau Momoko, Ko Tuu) prayed about the victory. He was turning round and round. Hence, the circle was a certain magic symbol, cf. Rapanui *taka*, *takataka* 'circle; to form a circle.' It was, in my opinion, the sign of the sun.

In the "*Apai*" text (Thomson 1891: 517-518) the following segments are presented:

(a) *Hata taka*. 'The sun (*taka*) is rising' (cf. Old Rapanui *hata* 'to rise,' Maori *whata* 'to elevate');
(b) *Mokomoko uri ua, Mokomoko tea, taka i a rangi*. (It is) the black Lizard (the god *Hiro*) as the rain, (it is) the white Lizard (= the clean sky), the sun (*taka*) is in the sky. (The translation is of mine.)

Examine several inscriptions associated with the read record. First and foremost consider the record on the Tahua tablet, see figure 14.

Ab 5-6: (a) 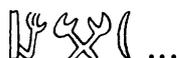 ...

(b) 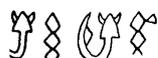

Figure 14.

Ab 5-6: (a) **4-15 20 3** *atua roa Ungu Hina* the great goddess 'The Crab – the Moon'...
(b) **21-17 3 21-17** *ko te Hina Kote* the White Moon (Rjabchikov 1987a: 364-365, figure 2, fragment 4).

A Rapanui myth about the creation of the Universe (Heyerdahl and Ferdon 1965: figure 147) sounds as follows:

*Ko Makemake he tuki ki roto i te vai i ava, i parokoroko*. (The god) *Makemake* copulated with the water, (fishes) *ava* (and) *paroko* (appeared).
*He tuki ki roto i te maea i ihoiho kiko nea*. (He) copulated with the stone, the fish *ihoiho* with the red (*nea = mea* because of the alternation of the sounds *m/n*) meat (appeared). (Cf. Rapanui *ihoiho* 'very hard stone,' too.)

*He tuki ki roto i te oone he topa ko Tive, ko Rarai a Hoa, ko te Nuahine ka Ungu a Rangi Kotekote*. (He) copulated with the earth, *Tive* (the god of the western wind and area), *Rarai a Hoa* (The Sun as the Master, Owner, Friend = the god *Tiki*, the description of the monument Hoa-Hakananaia), the Old Woman (the moon goddess *Hina*) as the Crab (Old Rapanui *ungu* 'crab') in the Clear Sky were born. (It is the translation of mine.)

The Crab was an image of the new moon when a solar eclipse, a horrible event, could happen. In the decoded inscription the quasi-syllables **21** *ko* and **17** *te* are presented twice: as the grammatical articles *ko te* (glyphs **21-17**) and as the word *kote* (*kotea*) 'white' (glyphs **21-17**) preserved in the reduplicated form (*kotekote*) in the manuscript.

Consider two parallel records on the same tablet, see figures 15 and 16.

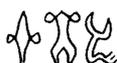

Figure 15.

Ab 8: **67-75 2** *Pikea Hina*. *Pikea (Uri)* as (the moon goddess) *Hina*.

Glyphs **67-75** read *piko* (to hide) also.

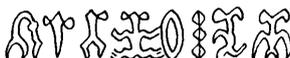

Figure 16.

Aa 8: **50/2** (= the inverted sign) **67-75 69 28-17 19-44** *Hina Uri – Pikea, Moko Ngotea kita*. (The goddess) *Hina Pikea (Uri)* (and the goddess) *Hiro Ngotea* [= *Paapaa Hiro* 'the Female (figurine) *Hiro*'] were the cause of death.

Glyph **50** *hi* was added to glyph **2** *Hina* to exclude the readings *mara* and *marama* (crescent). Again, glyphs **67-75** can be read as *piko* (to hide). Old Rapanui *ngotea* (< *\*ngo tea* 'the reservoir of the water during the bright sun') means 'to absorb,' cf. Rarotongan *ngotea* 'ditto,' Maori *ngongi* 'water' [< *\*ngo ngi*] and *ngongo* 'pool of water in the hollow of a rock, log, or tree' [< *\*ngo*].

Per a Rapanui myth (Métraux 1940: 260-261), the legendary king *Tuu-ko-Iho* carved the "portrait" of the female ghost *Paapa-ahiro* [*Paapaa Hiro* indeed] as a female wooden figurine (*moai paapaa*).

Glyphs **28-17** *Ngotea* are inscribed on the head of a wooden female figurine (*moai paapaa*) that is housed at the Peter the Great Museum of Anthropology and Ethnography (Kunstkammer), St. Petersburg (Rjabchikov 2009a: figure 37).

Consider the record on the same tablet, see figure 17. Here the words *ngotea* are written thrice.

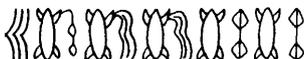

Figure 17.

Ab 6: **33 28 66 28 33 (=32) 28 33 (=32) 28-17 28-17** *Vai (ua): ngotea, ngo vai (ua), ngo vai (ua), ngotea, ngotea*. The water (rain): the water of reservoirs, the water of rains, the water of rains, the water of reservoirs, the water of reservoirs.

Consider the record on the Keiti (E) board, see figure 18.

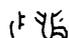

Figure 18.

Er 9: **46 20** *Naa Pikea.* (The deity) *Pikea* (*Uri*) is hidden.

Consider the record on the Small Washington board, see figure 19. I have reconstructed segment (c) relying upon a photo of the tablet. [2]

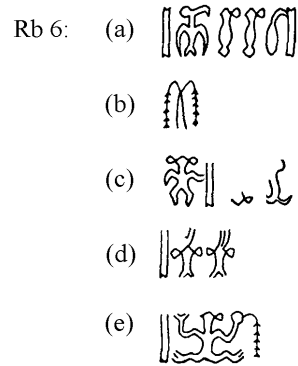

Figure 19 (corrected).

Rb 6: (a) **4-44 56-56-25-4** *Titaha: popohutu.* (The royal school) Hare Titaha: (write the word) *popohutu*.
(b) **24-24** *Ai, ai.* (Write) here, here (= again, again).
(c) **6-4 21 19** *Hati, ako: kuia*, Write, learn: (the word) *kuia*,
(d) **4 63-63** *tia* (or *ati*): *kapakapa*, write: (the word) *kapakapa*.
(e) **4 69 24** *Tia* (or *ati*), *moko ai.* Write (them) here (= again).

The word *pohutuhutu* is rendered in a variant of the Rapanui chant "*He timo te akoako*" (Barthel 1959: 168). *Kuia* and *kapakapa* are the basic words of that text (Rjabchikov 1993: 139-140, appendix 2, figure 5, fragment 75; 2012a: 565). In a Rapanui text known as the Creation Chant (Métraux 1940: 320-322) a personage's name contains the same term *pohutuhutu*. Fedorova (1978a: 334) compared it with Maori *pōhutu* 'surf.' I have disclosed such parallels: cf. Maori *pōhutu* 'to splash; surf,' Samoan *po* 'to slap' and *futiafu* [< *\*futi afu*] 'basin of a waterfall,' but also cf. Hawaiian *huku* [*hutu*] 'elevation.'

Consider the similar records on the Tahua and Great St. Petersburg tablets, see figure 20.

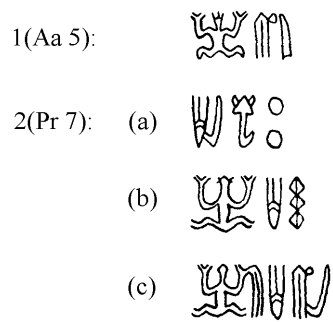

Figure 20.

1 (Aa 5): **69/1 26-5** *Moko, Tiki Matua* (the solar eclipse): the Lizard (*Hiro*; the new moon) (and) (the sun deity) *Tiki-Parent*
2 (Pr 7): (a) **1 5 35-21 139-139** *Tiki atua, Puoko Takataka*. The god *Tiki*, the Skull of the Round (the Sun).
(b) **69 1 17** *Moko, Tiki tea*. The Lizard (*Hiro*), the white *Tiki*. (The sky is clean.)
(c) **69 32 1 26-5** *Moko ua, Tiki-Matua*. The Lizard (*Hiro*) as the rain, *Tiki*-Parent.

The first record reports about a solar eclipse (Rjabchikov 1997c: 37).

---

[2] See that photo: <http://collections.si.edu/search/tag/tagDoc.htm?recordID=nmnhanthropology_8010183>.

The second record was a prayer to invoke rain. It is well known the skull was an incarnation of the god *Makemake* (Métraux 1940: 313). According to the legend "Hiva Kara Rere, the god of the rain" (Felbermayer 1971: 29-32), the priest *Rangi Taki* demanded that the god *Tiki* hide his face (*aringa*) before the beginning the rain during the drought.

### The Nature of the Rapanui God *Makemake*

Attempt to elucidate the origin of the cult of *Makemake* now. In this connection, consider the record on the Aruku-Kurenga board, see figure 21.

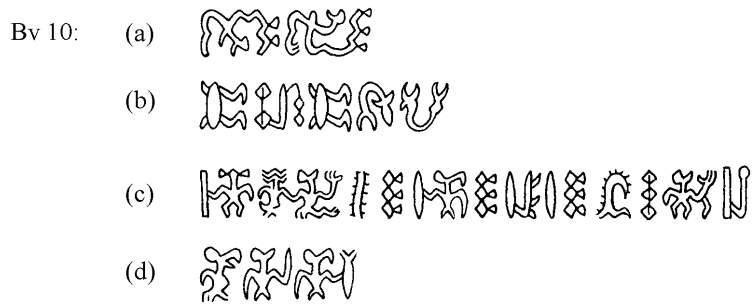

Figure 21.

Bv 10: (a) **50 31 52 31 52** *Hi: Make hiti, Make hiti.* The first pallor of dawn: the sun (the god *Makemake*) is rising, the sun (the god *Makemake*) is rising.
(b) **28-29 17 4 66 28-29 11-11** *Ngaru te Atua Tea, ngaru niuhiniuhi.* The White Deities (the great white sharks = *niuhi*) are floating, the numerous sharks *niuhi* are floating.
(c) **4-6 6-33: 19 4 17 30-44 17 30-51-30 17 14 17 6-15 4-4** *Tuha Haua: ku ati te Anakena, te Anakena, te Haua, te Hora Atuaatua.* The intervals of (the goddess) *Haua*: the month *Anakena* (this name is repeated twice), (the goddess) *Haua*, the month (season) *Hora*, (the 13th moon) *Atuaatua* (= the moon goddess *Haua*) are moving.
(d) **19-44 44-9** *Kita taniva.* The darkness (the sea waters) is the cause of death.

Per the Creation Chant (Métraux 1940: 320-322), the deities *Tingahae* (*Tinga hae* or *hoe* 'The paddle kills;' the object is the sign of authority) and *Parararara-hiku-tea* produced the shark *niuhi* (cf. *hiku tea* 'white tail,' the word \**para* is comparable with Tuamotuan *parāoa* 'whale' and *parata* 'species of shark;' cf. Maori *para* 'to shine clearly,' Tuamotuan *para* 'ripe,' Rapanui *pari* 'wave breaking on shore;' cf. also the name of the Old Peruvian god *Paryaqaqa*). *Makemake* played the role of the sun deity.
Consider the records on the Tahua, Small Washington and Aruku-Kurenga tablets, see figure 22.

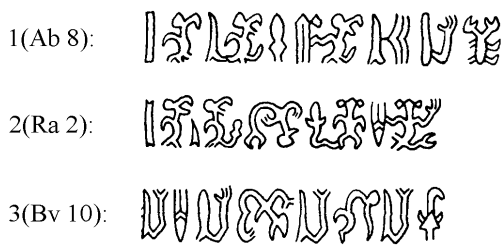

Figure 22.

1 (Ab 8): **4-19-5-19 73 26-19 5-33 4-15 20** *Tukituki e Maki-atua, atua roa Ungu.* (The god) *Makemake* copulated, the great goddess 'The Crab' (was born).
2 (Ra 2): **4-19-5-19 73 31 25 6 1 6 20var** *Tukituki e Make hua a Tiki, a Ungu.* (The god) *Makemake* copulated, the son *Tiki*, (the goddess) 'The Crab' (were born).
3 (Bv 10): **5-15 1 48-15 31 5-15 44b 5-15 35-21** *Atua roa Tiki uri Make, atua roa Tua, atua roa Puoko.* The great god *Tiki*, son of *Makemake*, the great god 'The Deep Water,' the great god 'The Skull.'

The name *Makemake* is written as the combination of two syllables (**26-19** *ma-ki*) and the ideogram (**31**). The first fragment is presented at the end of the Tahua inscription that was a manual in the *rongorongo* school of king *Kai Makoi* the First. The second fragment was a copy of that text with an addition in a manual from the school of king *Nga Ara*. In the third record Old Rapanui *uri* 'son' is put down, cf. not only Rapanui *ure* 'penis; son,' *huri* 'new shoot (of banana),' Maori *uri* 'offspring; descendant,' but also Quechua *churi* 'son.' *Makemake* was an outside god for the tribe Miru at first, and among the western tribes of Easter Island that deity was provided with the features of their principal sea god *Tangaroa*. According to Thomson (1891: 482), *Makemake* (*Meke-Meke*) was the great spirit of the sea.

Glyph **31** *Maki*, *Make* represents a zoomorphic (feline) spirit. I distinguished Quechua *puma* 'puma' in the Rapanui place name *Pumakari* earlier (Rjabchikov 1997d). One can suggest that glyph **31** depicts that beast.

(1) Consider the drawing on a wooden bowl from the former Incan empire dated to the colonial times (Stingl 1986: 143, the lower figure). The puma is represented there. On its body there are 26 rhombi and one **X**-like sign (on its head). I suggest that the rhombi (cf. Quechua *punchai* 'day,' *inti* 'the sun') are the symbols of days of a lunar month as well as of the sun. See above about 26 dots on a Rapanui panel.

Per Zuidema (1985: 184, 187), the puma in the Incan empire was a certain symbol connected with December and June, both months of solstices during the year. The skin of that animal was associated with the ripe harvest. I have arrived at the conclusion that the Old Peruvian **X**-like sign denotes the terms 'harvest' and 'to grow,' 'growth' (cf. Quechua *allay*, *pallay*, *kuxichu* 'harvest,' and *wiñay*, *wichay* 'to grow').

(2) Consider an Old Peruvian textile pattern (Galich 1990: 378, figure). Here a standing man is depicted. I have interpreted the design. Between his feet on the ground a symbol (*tiqsi*) resembling the vulva is shown. A rhombus (*inti*) is seen above it. On the belly a rectangle with signs of waves (*qucha*) is presented. Above that symbol an octahedron (*wira*) is presented. So, the name of this god is *Tiqsi Inti Wiraqucha*. It was the combination of two Incan deities: the sun god *Inti* and the god-creator *Cun* (*Con*) *Tiqsi* (*Tiksi*) *Wiraqucha* (*Wiraqocha*). Here the Quechua terminology has been of our main interest: cf. Quechua *kunya* 'flame,' *kunununu* 'thunder,' *tiqsi*, *ti'qsi* 'prime cause; basis, foundation, origin, source,' *wira* 'fat' and *qucha* 'lake.' Cf. also Aymara *tik'i* 'origin; basis' and *qota*, *q'ota* 'lake.'

(3) Consider a Moche vessel representing a warrior; on his face the penis and vulva, birds, animals and other signs are depicted (Scher 2010: 280-282, 439, figure 6.31; 442, figure 6.37). In addition, I propose the interpretation of four red rays on the body of that warrior. No doubt it is the designation of the dawn. In this case, that personage was the sun god having a characteristic of the god-creator. It was a prototype of the united deity with the traits of the Old Peruvian gods *Inti* and *Cun Tiqsi Wiraqucha*. It should be pointed out that some warrior figures modified in the same style in Moche ceramics wear a feline circlet (Scher 2010: 280-281, 441, figures 6.35 and 6.36; cf. Heyerdahl 1976: figure 32).

(4) Consider the depiction on another Moche vessel: a double-headed bird motif has been distinguished (Scher 2010: 265, 369, figure 5.94). This design is comparable, in my opinion, with double-headed frigate bird motif in the Easter Island rock art and script. Such glyphs in some cases are associated with the cult of the god *Makemake* (Rjabchikov 2012a: 565-566, figure 1).

On Easter Island the union of the Old Andean and Polynesian creation myths occurred. In the version of the myth that is known from the works of the Norwegian archaeological expedition and translated by me above, *Makemake* did not separate the Sky from the Earth as in the stories of Polynesia, but copulated with different substances (water, stone, earth). In records 1 and 2 of figure 22 the process of the copulation is described itself, without the participation of the earth. The Rapanui term *tukituki* (to copulate) therefore is a prominent feature of the god *Makemake* (*Make-Make*).

In the Incan mythology the Quechua words *Cuniraya Huiracocha runakamaq, pachakamaq* mean '*Cuniraya Huiracocha*, the human being maker, the earth maker' (Taylor 2008: 22-23). The archaic name *Cuniraya Huiracocha* = *Cun Iraya* (cf. Quechua *illa*, *rayu* 'lightning' because of the alternation of the sounds r/l) *Wiraqucha*, the same Incan and pre-Incan god-creator. Hence his basic function was the creation of humans and this world. Quechua *kamaq* means 'maker.' The Tuamotuan term *makemake* refers to male sexual life, cf. also Tuamotuan *make* 'innumerable.' It was a borrowing from Quechua. (The Old Peruvians visited not only Easter Island, but some other islands of the Pacific.) It is likely that the archaic Rapanui word *makemake* meant the same. Because the god *Makemake* (Maker, Creator) was the sun deity, its name was interpreted by the Easter Islanders in the solar terms (*Maa*, *Maa-ke*, *Maa-ki*) later.

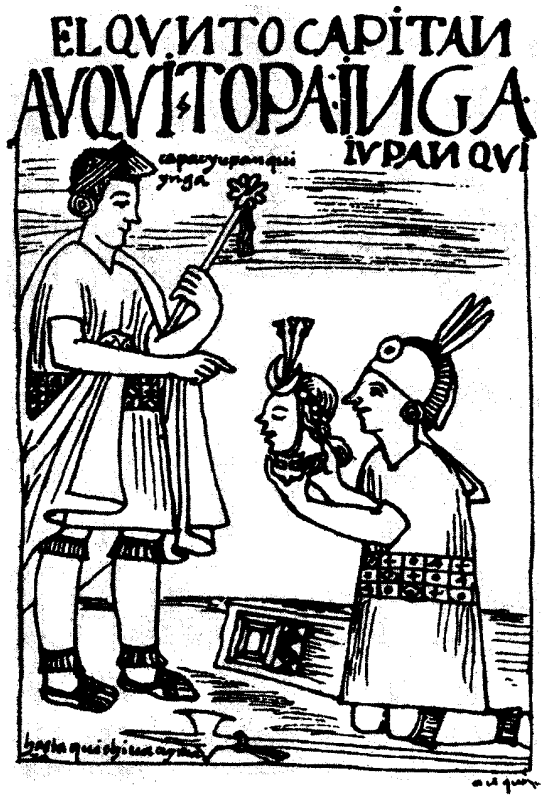

Figure 23.

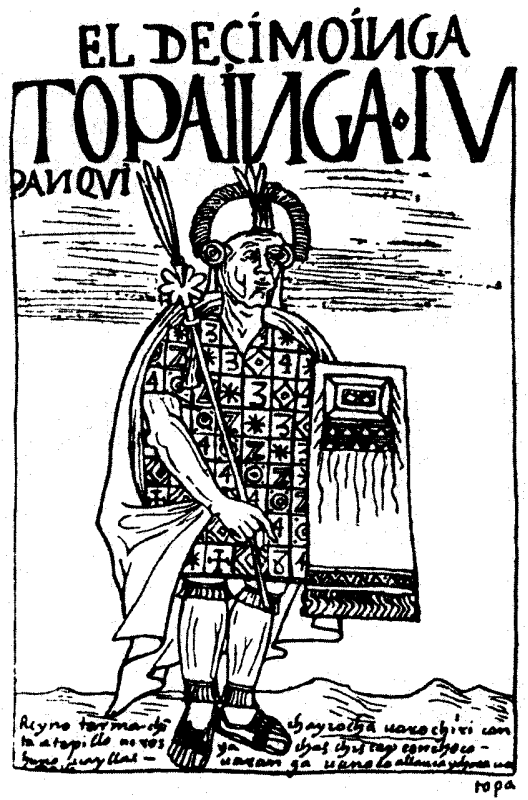

Figure 24.

(A) 4 · ◇ ✢

(B) · 4 ✢

(C) 𝒵 4

(D) ⊙

(E) 𝒵 ✺

Figure 25.

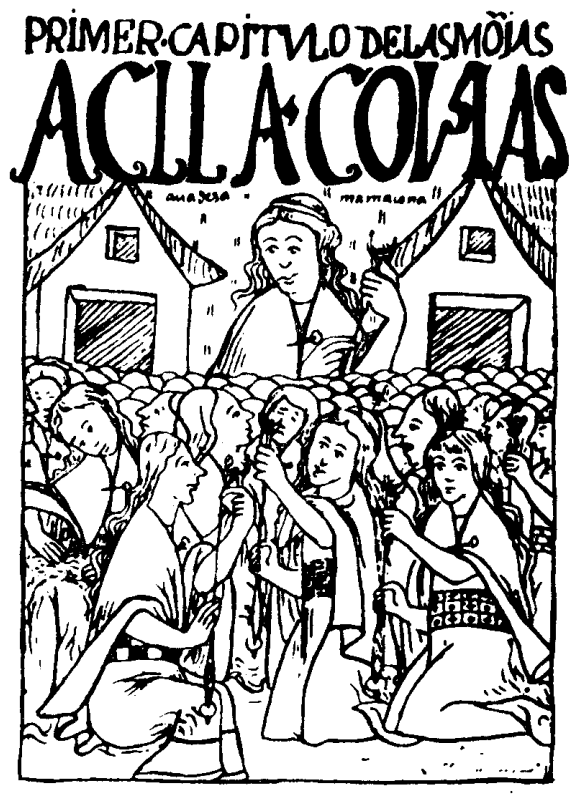

Figure 26.

• ✢

Figure 27.

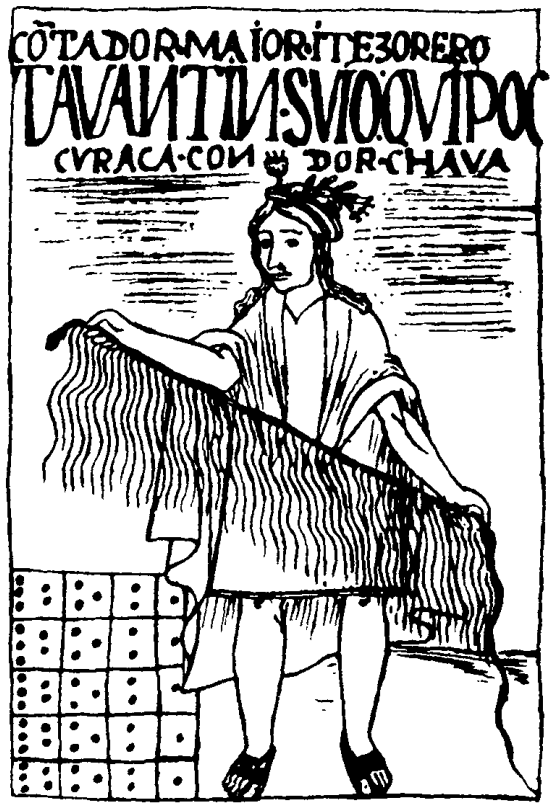

Figure 28.

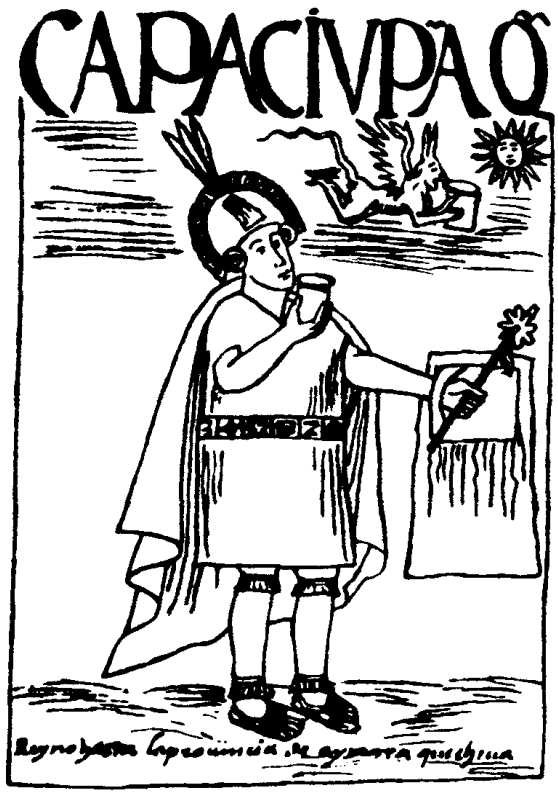

Figure 29.

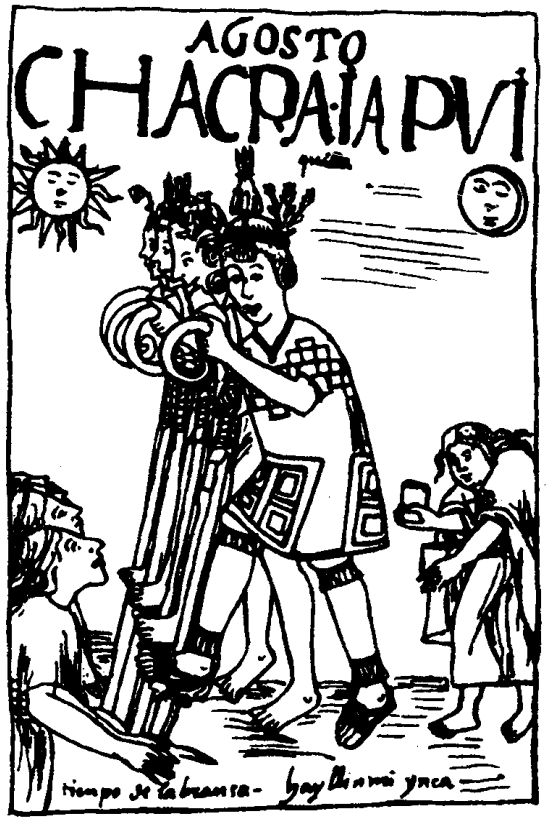

Figure 30.

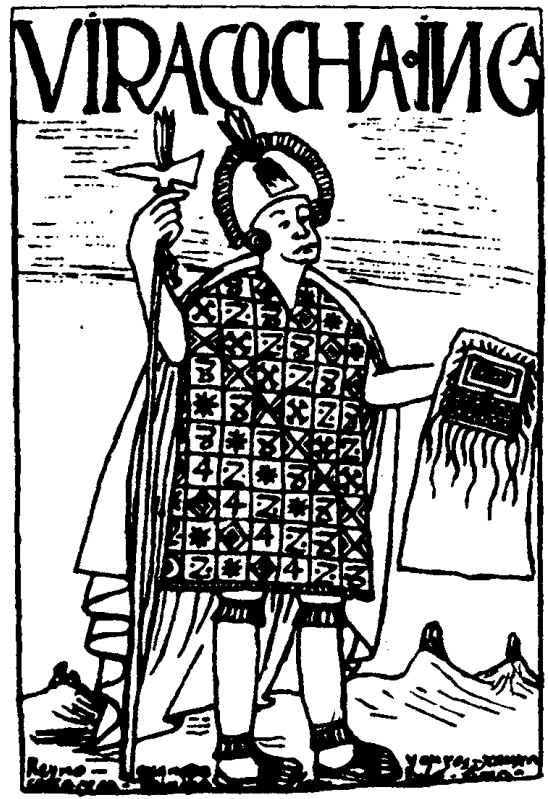

Figure 31.

(A) ↶ ◊ ☀

(B) ✕

Figure 32.

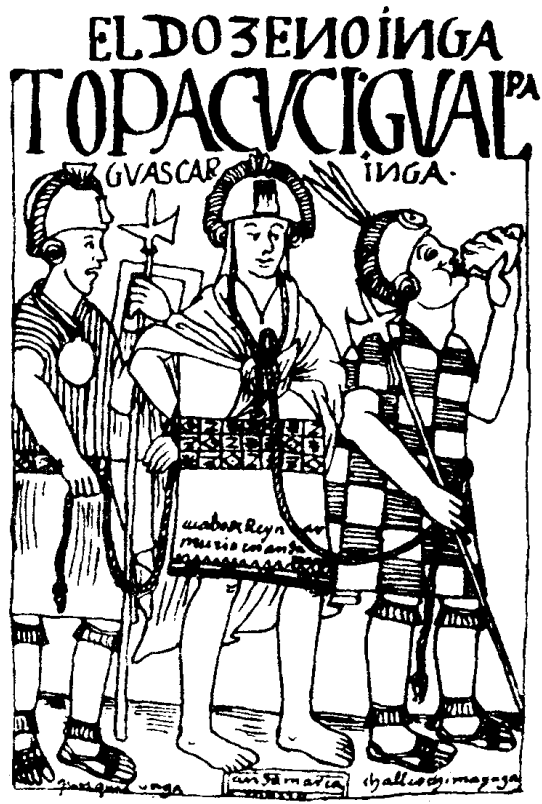

Figure 33.

(A) Z ◊

(B) Z ☀

Figure 34.

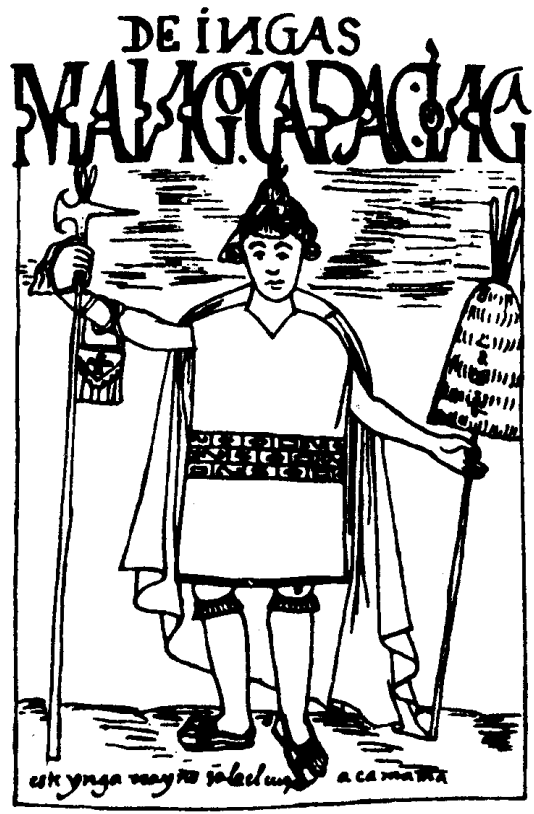

Figure 35.

(A) 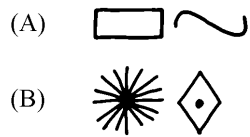

(B)

Figure 36.

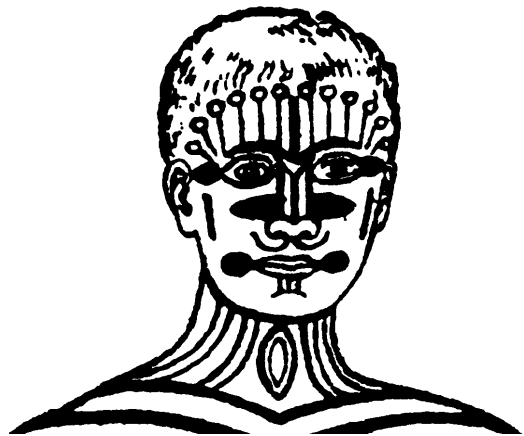

Figure 37.

# The Incan Linear Script. The Parallels in the Culture of Easter Island

Except for the *quipu* (knots on strings), the Old Peruvians used the other script where signs were certain ideograms (realistic and abstract pictures). Let us consider several examples. Here I investigate drawings from the ancient book "The First New Chronicle and Good Government" (1600 to 1615 A.D.) by Guaman Poma de Ayala, see figures 23, 24, 26, 28, 29, 30, 31, 33 and 35. I have gathered some glyphs from these drawings in figures 25, 27, 32, 34 and 36. It is the continuation of my studied of the Old Peruvian writing system (Rjabchikov 1994: 44-56). It is important to note that I have already analysed several symbols of the Old Peruvian art (writing), see above. The Old Andean script could be one of sources of the *rongorongo* script of Easter Island.

1. In figure 23 a ruler of the Incan empire is depicted. His attire was adorned with signs shown in the diagonal order. Nevertheless, having read them from left to right, one can pick over some segments of the local writing.

According to Silverblatt (1987: 97), an Inca told secret words to the sun (the god *Inti*) during sacrifices (*qhapaq hucha*): "Receive these chosen ones for your service." The sacrificial ceremony was called *qhapaq hucha* (*capac hucha*) (Abercrombie 1998: 4). Quechua *qhapaq* means 'mighty; sacred; grand,' *hucha* 'victim, sacrifice,' cf. *arpa* 'victim; offering' as well.

Quechua *chaskiy*, *abiniy*, *uya-kuy* mean 'to receive', *apay* 'to take.' The +-like glyph denotes the words 'to receive; to take.' The **4**-like glyph denotes the words 'sacrifice, victim, offering' (*hucha*; *arpa*). The rhombus denotes the sun (*inti*) and the day (*punchai*).

Per that drawing (figure 23), the Inca received the head of an enemy. Fragment (A) of figure 25 means 'Receive the victim for the first sun!' Fragment (B) means 'Receive the first victim!'

2. In figure 24 the emperor *Tupac Yupanqui* (1471 – 1493 A.D.) is depicted. It is well known that his crew (and perhaps he himself) voyaged to Polynesia (Heyerdahl 1968: 45ff). The Rapanui tribe Tupa-Hotu could be named after that Inca (Rjabchikov 1987b). Fragment (C) means '(human) sacrifices' (*qhapaq hucha*). Fragment (D) means 'the sun' (*inti*) or 'the god *Inti*', and the four dots surrounding the sun symbol designate the four cardinal directions. Fragment (E) means 'the great son' (*hatun churi*), and this formula proclaimed the divine origin of this Inca.

3. In figure 26 the Old Peruvian "brides of the sun" are depicted. On the garments of one of such girls certain signs are seen. The brief record is presented in figure 27. It means 'Receive (me) first!' These words were said to the sun deity *Inti* and to an Inca, as to his representative on the earth.

On the other hand, Rapanui children called *neru* lived separately in a special cavern called *Ana hue neru* according to the legend "Children in Isolation" (Englert 2002: 166-169; see also Lee 1992: 47). The term *neru* came from the expression *\*naa rua* 'hidden.' The name *Ana hue neru* means 'The cave (where) *neru* were gathered (*hue*).' Another name of their shelter, *Ana o keke*, located on the Poike peninsula signifies 'The cavern of the sunset (to save the skin from tanning).' Rapanui *ngangi* means 'to be a virgin' < *\*nga ngii*, cf. Rapanui *nga* 'the grammatical article (Plural) of nouns;' *ngii* 'glare of the sun,' and *ngiingii* 'burning.' So, the term *ngangi* means '(that who belongs to) the bright sun' literally. Rapanui *nire* 'virgin' derived, in my opinion, from the expression *\*ngii raa* '(that who belongs to) the sun.'

4. In figure 28 a man holding a *quipu* is shown. In the squares dots are presented. They are number 1 and three prime numbers: 2, 3 and 5. The numbers were chosen so that 1+2=3 and 2+3=5. Perhaps that man was an Old Peruvian mathematician and astronomer.

In the Rapanui rock art dots (cupules) were wide used to count numbers (months, nights/days, perhaps murdered persons and so forth).

5. In figure 29 an Inca who prayed to *Inti* is shown. On his belt solar signs are seen. The devil (a Christian image) is represented together with the sun. The observations of the sun were conducted in the Incan state (Stingl 1986: 206). On Easter Island the "solar stones" at the ceremonial village of Orongo were the real observatory to determine the days of solstices and equinoxes (Ferdon 1961; 1988).

6. In figure 30 an Inca and others digging a field are presented. His attire was adorned with at least two specific signs (a square divided into four parts). Such a sign means 'field' and 'earth' (cf. Quechua *pacha* 'earth,' *pampa* 'field,' *allpa* 'cultivated soil' < *\*pa*).

In the Rapanui rock art several tribes (their territories) are designated as united rectangles (Rjabchikov 2001: 216).

7. In figure 31 an Inca is presented. His attire was adorned with several signs. Two records are collected in figure 32. Record (A) means 'The son of the mighty sun god *Inti*.' Record (B) means 'Harvest.'

8. In figure 33 the Inca *Huascar* is represented. It is common knowledge that *Atahualpa* (*Atawallpa*) overthrew his brother (Brundage 1985: 296). It is very likely that the records on the king's attire confirmed his royal power. Two inscriptions are collected in figure 34. Record (A) means 'The great sun god *Inti*' (*Hatun Inti*). Record (B) means 'The great son' (*Hatun churi*).

9. The first legendary Inca called *Manqu Qhapaq* (*Manco Capac*) is represented in figure 35. On his attire two inscriptions are found, see figure 36. Record (A) reads *Manqa* (*Manqu*) *Qhapac*, it is the exact name of the Inca. The first sign is a rectangle, the designation of something put into some volume. Cf. Aymara *manq'a* 'within.' One can suspect that the Inca and some others spoke the Old Quechua language, the forerunner of Quechua and Aymara. Record (B) reads *Churi Inti*. This expression meaning 'The Child Sun' or 'Daylight' was an aspect of the Incan sun god *Inti* (Conrad and Demarest 1984: 107).

### The Kings of Easter Islands as Images of the Sun

King *Hotu-Matua* of the western tribes received, as could be expected, some of the features of the sun deity under the influence of the ideology of the eastern tribe Tupa-Hotu (Rjabchikov 1995; 2009b). The name *Tupa-Aringa-Anga* of the last Rapanui king from the Tupa-Hotu tribe (Hanau Eepe) consists of the ethnicon *Tupa* (*Tupa-Hotu*) and the name *Aringa-(H)anga* 'The Face (the sun) is moving.'

Consider the "portrait" of king *Nga Ara* represented in figure 37 (Routledge 1998: 219, figure 88; the interpretation in Rjabchikov 2009a: figure 20; 2012a: 567). On the forehead of the monarch 12 small rounds are depicted. On his nose glyph **1** *Tiki* (the name of the sun deity) is revealed. It is apparent that those 12 rounds denote the sun during the whole year (12 months). I suppose that such signs were tattoos of the eastern kings (Hanau Eepe; Tupa-Hotu) at first, and later the kings from the Miru group (Hanau Momoko) became to decorate their faces with the similar patterns.

As a parallel, consider the face of the central personage (the sun deity *Inti*, the god-creator *Viraqucha*) of the Gate of the Sun at Tiwanaku (see Heyerdahl 1976: figure 44). Count the signs of the sun attached to all the sides with the exception of the chin (the lower part of the head): I have received the 12 signs. (The Rapanui glyphs **39**, **115** *raa*, *taka* 'the sun' resemble them.) The other signs added to the god's head are feline's heads (*puma* etc.) and the rectangle divided into some parts (*Pacha* < *Pacha Kamaq*). The deity holds in his hands spears tipped with condor's heads (cf. the name *Con* or *Cun*). The image of that god could be a prototype of the tattoos of some Rapanui kings (chiefs).

### The Old Andean-Polynesian Lexical Parallels: the Divergence or Convergence?

If we agree that the image of the Old Andean god *Con (Cun) Tiqsi* (*Tiki*) *Wiraqucha* was once introduced into the Rapanui (Polynesian) culture, the term *Hanau Eepe* will be understandable well. The component *Wira-* of the name means 'Fat,' and the Rapanui term *Hanau Eepe* means 'The fat (= rich) people.'

One of the platforms on Easter Island is called *Tumuheipara* (Thomson 1891: 504). It reads *Tumu hei para*, where Old Rapanui *tumu* (*tuma*) signifies 'foundation, basis.' The unclear term *hei* is cognate with Maori *heihei* 'storm.' One can therefore presume that here *Para* is a part of the name of the Old Peruvian storm god *Paryaqaqa*. On a panel (see Lee 1992: 52-53, figure 4.11) at the place Papa te Kena located nearby we see the mask with rays as well as the round with the dot in the centre both closely related to birds. I conceive that they are the images of the god-creator (*Cun Tiqsi Wiraqucha*, *Paryaqaqa*).

The Rapanui stone towers *tupa* are compared with Old Andean burial stone towers *chulpa* (Heyerdahl 1968: 194; Langdon 1994: 77). One can try to interpret the latter term: 'The hidden or sectet (place),' cf. Quechua *ch'ulla* 'sole, single' and *pakay* 'to hide.' The name of the Rapanui feast *Paina* associated with the harvesting can be explained from Maori *paina* 'to warm oneself.' On the other hand, let us remember Quechua *apay* 'to take.' Then the term *Paina* (*Painga*) means '(The offerings) took (by the sun god).' Rapanui *hu* is a rare grammatical article of nouns (Englert 1948: 442). I suggest that it is cognate with Quechua *huh* 'one,' cf. Rapanui *etahi* (*e tahi*) 'one' standing before some nouns.

A number of lexical parallel between the Quechua and Polynesian languages have been found, and the source of borrowings was the first language. Cf. Quechua *kipu* 'type of the script: knots on strings' and Rapanui *kupu* 'text; time;' Quechua *raphi* 'leaf' and Rapanui *raupa* 'large leaf' (the addition *pa* is seen after the standard root *rau* 'leaf'); Quechua *atiq* 'victor' and Rapanui *titikanga* (< *tika*) 'authority;' Quechua *kanchay* 'to build around' and Rapanui *kato* 'to build;' Quechua *pakay* 'to hide' and Rapanui *piko* 'ditto;' Quechua *kanay* 'to burn,' *k'ancha* 'shine' and Tuamotuan *kana* 'to shine;' Quechua *killa* 'the moon,' *killay* 'to shine (of the moon)' and Rapanui *kii* 'ditto;' Quechua *wañuy* 'death' and Rapanui *vana-vana* 'headdress used during a combat;' Quechua *wawa* 'child; baby' and Rapanui *vovo* 'daughter; girl;' Quechua *ñuñuy* 'to breastfeed' and Old Rapanui *nua* 'mother' (but cf. also Aynu *unu* 'ditto.')

Thus, at least some similar Old Andean and Polynesian (Austronesian) forms could appear independently from each other, when Old Asiatic people and their descendants went to North and South America in the Palaeolithic era, and other Asiatic people and their descendants swam in the Pacific several thousand years ago.

Aymara *p'eqe* 'head' and Rapanui *puoko* 'skull; head' (Samoan *ulupo'o* 'skull' < *uru poko*), Quechua *apu* 'lord' and Samoan *tupu* (< *tu pu*) 'king' (Maori *pu* 'ditto,' Old Rapanui *pu*, *pua* 'ruler; king'), Quechua *ninay* 'to set fire to' (the root *ni*) and Rapanui *ngii* 'glare of the sun' (Samoan *tugia* 'to set fire to' < *tu gi*), if such parallels are valid, can be explicated from certain Palaeolithic forms: *pa- 'bearer of benefits; protector; saviour,' cf. Turkic *bash* 'head,' Aynu *syapa* 'ditto;' *pi-/pu- 'great; strong; master,' cf. Kongo *buma* 'great,' *ekabu* 'very strong man,' Sanskrit *bhuri* 'grand; important;' *ni-/nu- 'fire,' cf. Kongo *anyuyi* 'burning fiercely,' Sanskrit *agni* 'fire.' A great number of such lexical parallels with the glossary of the language of the Palaeolithic people as well as the analysis of their sign system are presented in earlier works of the author (Rjabchikov 2006; 2007a; 2007b).

## The Problem of the Origin of the Mamari Tablet

Rapanui *ro* (in the combinations *ro ai*, *ro atu ai*, *ro mai ai*, *ro ana*) was an archaic verbal particle following after the word (Fedorova 1978b: 73). In my opinion, the Rapanui particles *ro ai* are cognate with Quechua *raya* (the verbal affix of the Continuous Tense). The Polynesian particles *atu* (there), *mai* (from), *ana* (numerous) were introduced into the Quechua grammatical structure *raya* > *ro ai* later.

Let us examine two sentences from the Rapanui legend "*Ko Ko Rou o Rongo*" (Heyerdahl and Ferdon 1965: figure 165-167; Fedorova 1978a: 206-208):

**Ku** *tuu* **ro ai** ... (They once) came...
**Ku** *rangi* **ro atu ai** *te tahi nga poki: "Hai pakahera tatoua matau!"* One of the young men shouted: "We shall fight with calabash-headed spears!"

In figure 1, segment (46) the text with the similar structure is rendered, see above: **Ku** *hiti*, **ku** *hea* **ro atu**... The particles *ku* denote the Past Tense.

Let us examine two fragments of texts where the verbal particles *ro atu* without the final particle *ai* are found; they are taken from the Rapanui legends "Insult Songs" and "The Story of *Makita* and *Roke Aua*" (Englert 2002: 184-185, 236-239):

*E ei* **ro atu** *hoki au mo ou*… I shall sing one against you, too…
*E tu'u* **ro atu** *au.* I shall catch you. (I shall come to you literally.)

Fedorova (1978a: 314-315) correctly translated the beginning of the chant "*He timo te akoako*" from the manuscripts discovered by the Norwegian archaeological expedition (Heyerdahl and Ferdon 1965: figure 127):

| | |
|---|---|
| *He timo te akoako,* | A pupil is carving (the glyphs), |
| *he akoako tena,* | (he) is learning them, |
| *e te tuu,* | the direction, |
| *e te taha,* | the turn: |
| *e te kuia,* | the bird *kuia*, |
| *e te kapakapa…* | the bird *kapakapa*… |

The pupils toiled at the text about initiations in the royal *rongorongo* school. The words e *te manu vae punaka, e te manu vae e ha* (the bird with the legs of a chicken, the bird with four legs = rapid legs; it is my own translation) from the chant describe, in my opinion, boys (future valiant warriors) during those rites in December (the month *Koro*; perhaps, the ceremonies lasted two days, see figure 1).

A more archaic version of the chant was presented on the "Tablette échancrée" (D), see figure 38.

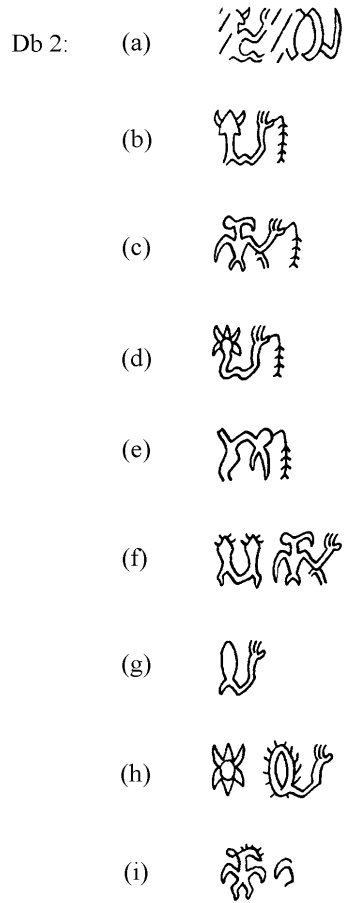

Figure 38.

Db 2 (= Dr 2?): (a) **21 69 57 5** *Ako moko, tara, tia,* A pupil is carving,
(b) **21 15-24** *ako ro ai,* (he) is learning (the glyphs),
(c) **44 15-24** *taha ro ai,* (he) is turning (the tablet),
(d) **7 15-24** *tuu ro ai:* (he) is directing (the glyphs):
(e) **19 24** *kuia, ai,* (the glyph) *kuia*, (write it) here (=again),
(f) **63-63 44 15** *kapakapa taha (= manu) ro,* and (the glyph) *kapakapa* BIRD (the determinative),
(g) **12 15** *ika ro,* and (the glyph) *hika*,
(h) **7 14 15** *tuu Hau ro,* and (the glyphs) 'the star Antares,'
(i) **49-49** *(Ariki) Mau-(Ariki) Mau.* (the glyphs) 'the star Antares.'

The brief glossary: only Thomson (1891: 548) is the source for Rapanui *kuia* 'boobies (birds),' but other sources give Rapanui *kena* 'ditto.' I suggest that the form *kuia* is a phonetic variant for *kia*, cf. Rapanui *kiakia* 'white tern.' According to the Creation Chant (Métraux 1940: 320-322), the deities *Kuhikia* and *Rupe roa* (the great Pigeon) produced *turi* (the sea-gull). The form *kuhikia = kui-kia = kiakia*, hence *kui, kuia = kia, kiakia* (white tern) due to the alternation of the sounds *u/i*. The term *kapakapa* is cognate with Rapanui *kakapa* (< *kapa) 'certain sea bird.' Old Rapanui *hika* 'boy' is cognate with Maori *hika* 'term of address to young persons of both sexes.' Old Rapanui *hau* means 'king,' cf. Tahitian *fau* 'ditto.'

In the deciphered record the verbal particles *ro ai* are written thrice. Several variants of the chant (Routledge 1914-1915) contain such a segment: (*e te*) *kotiro* that is in reality (*e te*) *koti ro* 'and (other) glyphs (carvings literally),' cf. Rapanui *kokoti* 'to cut,' *kotikoti* 'to tear,' Mangarevan *kotikoti* 'to cut' and Marquesan *koti* 'ditto.' One variant of the chant (Routledge 1914-1915) as the text of a *rongorongo* board contains these words: *e te hau topa mai te ragni*. I have translated that expression as '(the star) *Hau* born from the sky.' It was the description of the first rising of the red star Antares (α Scorpii) before dawn. In the Maori (New Zealand) astronomy a star which marks the sixth month (November-December) of the local calendar is called *Ariki-rangi* 'King in the sky literally' (Best 1922: 31).

Consider the record on the Small St. Petersburg tablet, see figure 39.

Qv 3: 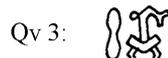

Figure 39.

Qv 3: **65 49** *RANGI (Ariki) Mau*. SKY (the determinative) Antares.

Let us run the computer program. In 1640 A.D. the first rising of Antares before dawn occurred on December 16. In 1860 A.D. that event fell out on December 19. Consequently, this star was a good herald of the summer solstice for the Easter Islanders.

Two scenes in the rock art – a lunar calendar and a story of fishery mentioned above – were the sources for the Mamari calendar record (see figure 1). Both rock drawings are located in the area of the ceremonial platform Ahu Raai. It was the territory of the Hanau Eepe (the Tupa-Hotu tribe). So, these petroglyphs can be dated no later than 1682 A.D.

## Conclusions

The beginning of the calendar record inscribed on the Mamari tablet has been dated to the day of the summer solstice of December 20, 1680 A.D. The moon was not visible earlier at night. Because of a possible solar eclipse it was a perilous day, a precursor of the future misfortunes: the motion of Halley's Comet of 1682 A.D. and the rebellion of the western tribes. The new data about the watchings of the star Antares have been obtained, too.